\documentclass[final]{IEEEtran}
\usepackage[T1]{fontenc}
\usepackage{times}
\usepackage{graphicx}
\usepackage{amssymb}
\RequirePackage[hyphens]{url}
\usepackage{balance}
\usepackage{dblfloatfix}

\begin{document}

\title{Turning Internet of Things(IoT) into Internet of Vulnerabilities (IoV) : IoT Botnets \thanks{\textbf{Disclaimer:} The views expressed here are solely those of the author in his private capacity and do not in any way represent the views of the Munich RE, or any other entity of the Munich RE Group}
}

\markboth{Turning IoT into IoV : IoT Botnets }        
{Kishore Angrishi}

\author{Kishore Angrishi


\thanks{K. Angrishi is with Munich RE, K\"oniginstrasse 107, 80802 Munich e-mail: KAngrishi@munichre.com.}
\thanks{The author like to thank Jo M\"uller, Carsten Topsch, Sebastian Wolf and Wilhelm Reeb for their time and feedback in both initiation into work on IoT and in the preparation of this contribution}
}


\maketitle

\begin{abstract}
Internet of Things (IoT) is the next big evolutionary step in the world of internet. The main intention behind the IoT is to enable safer living and risk mitigation on different levels of life. With the advent of IoT botnets, the view towards IoT devices has changed from enabler of enhanced living into Internet of vulnerabilities for cyber criminals. IoT botnets has exposed two different glaring issues, 1) A large number of IoT devices are accessible over public Internet. 2) Security (if considered at all) is often an afterthought in the architecture of many wide spread IoT devices. In this article, we briefly outline the anatomy of the IoT botnets and their basic mode of operations. Some of the major DDoS incidents using IoT botnets in recent times along with the corresponding exploited vulnerabilities will be discussed. We also provide remedies and recommendations to mitigate IoT related cyber risks and briefly illustrate the importance of cyber insurance in the modern connected world.
\begin{IEEEkeywords}
DDoS,  IoT, IoT Botnets, Mirai Botnet, Cyber Insurance, Security
\end{IEEEkeywords}
\IEEEpeerreviewmaketitle
\end{abstract}

\section{Introduction}
\label{sec:intro}
\begin{figure}
\begin{center}
 \includegraphics[width=0.48\textwidth]{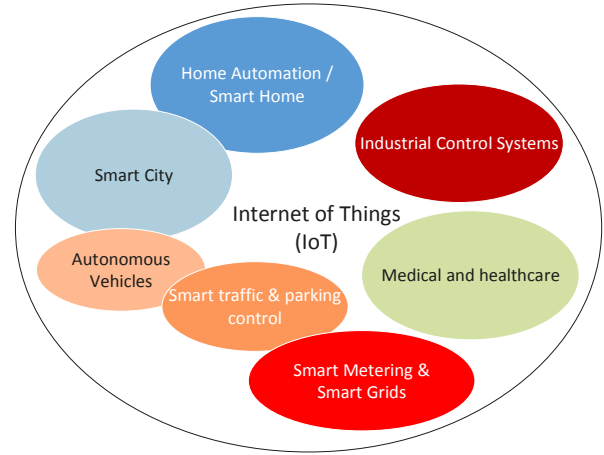}
\end{center} 
\caption{Internet of Things (IoT)}
\label{fig:IoT}       
\end{figure}

\begin{figure}
\begin{center}
  \includegraphics[width=0.48\textwidth]{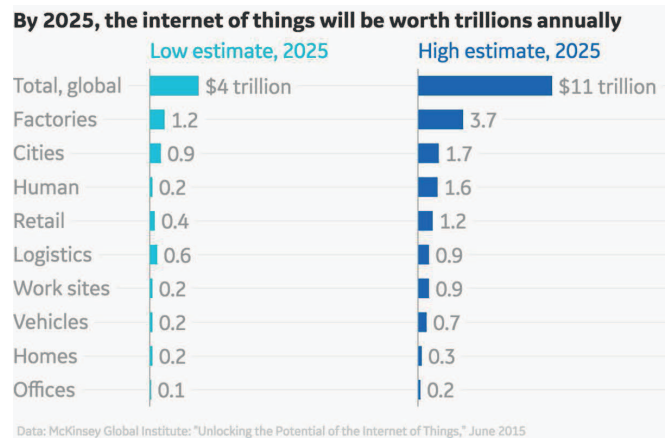}
\end{center}
\caption{The worth of IoT by 2025}
\label{fig:IoTWorth}       
\end{figure}

The Internet of Things (IoT) is key in the digital world of connected living. The futuristic appeal to make life bit more enjoyable in a hectic day-to-day routine is enticing to many. For example, the idea of refrigerators monitoring their contents and send orders directly to the retailers when the milk is running out or ordering Sunday morning bread from your bed with a voice or gesture command to an intelligent assistants like Amazon Alexa or Apple Siri or Google Assistant is appealing. With the advent of smart phones, smart television, and more smart devices like Amazon echo with Alexa or Google Home, most of the ideas stated above are not part of some science fiction dream anymore but rather becoming a reality right now. The IoT devices have a wide range of applications (see Fig: \ref{fig:IoT}) especially in  home automation (smart home), healthcare, smart energy solutions, autonomous connected vehicles and extremely complicated industrial control systems. According to a study \cite{IoTStudy} we have now (2016) have 9 billion smart devices (excluding smart phones, tablets and computers) and is anticipated to grow to 28.1 billion by 2020. By 2025, the value of the internet of things will be trillions annually (see Fig: \ref{fig:IoTWorth}) \cite{IoTWorth}. 

It is important to understand that these smart devices cannot be seen as specialized devices with intelligence built-in but rather as computers which does specialized jobs. For examples, a smart phone can be seen as a computer that makes phone calls or a refrigerator is a computer that keeps things cold. These specialized computers are often run by powerful microprocessors just as much as desktop, laptop or tablet computers and are well connected with each other, either inside a private network or over the public Internet. The crucial distinction with these specialized computers is that, IoT devices are often designed with poor security or even none at all. Internet is already very complex to secure, with additional 9+ billion insecure IoT devices, the task has become more difficult. In the next section, a brief introduction of simplified Internet is given to understand the vulnerabilities used by IoT botnets to launch attacks on clueless victims in Internet.

\section{Introduction to Simplified Internet}
\label{sec:internet}
\begin{figure}
 
 \begin{center}
	\includegraphics[width=0.48\textwidth]{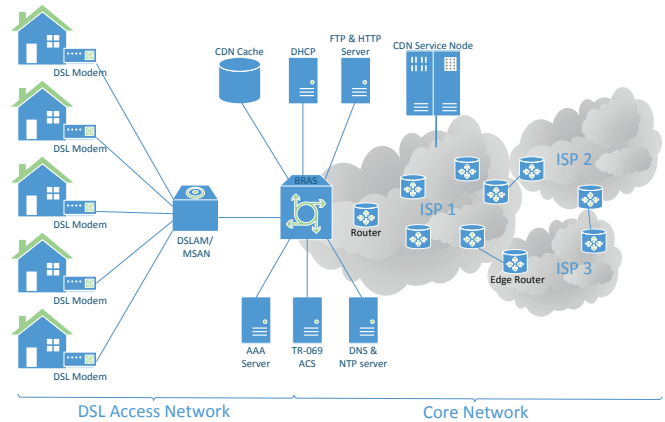}
 \end{center}
\caption{The DSL access network together with core network of ISPs}
\label{fig:DSLAccess}       
\end{figure}

\begin{figure}
\begin{center}
  \includegraphics[width=0.48\textwidth]{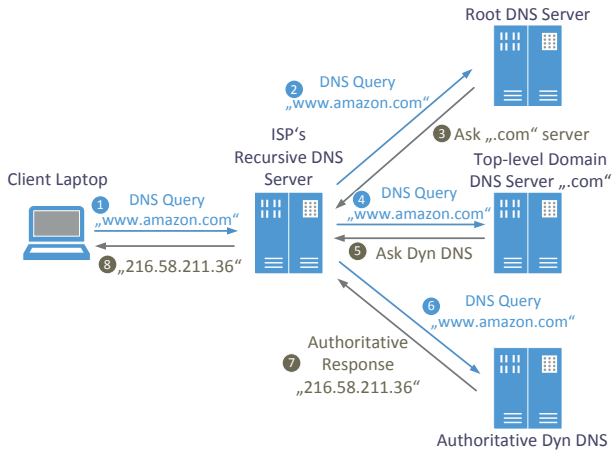}
\end{center}
\caption{The DNS query processing}
\label{fig:DNSOperations}       
\end{figure}

Internet is the network of networks. It can be broadly divided into two parts, namely, access network and core network. There are arrays of technologies involved in enabling the normal functioning of both access and core networks. Since IoT devices are mostly end customer devices deployed at edge of access networks and IoT botnets often exploit the vulnerabilities present at the interface between access networks and core network, we limit our discussion on internet to access networks especially digital subscriber line (DSL) access network (see Figure \ref{fig:DSLAccess}) along with the TCP/IP protocol suite. The first entity on the DSL access network is DSL modem located at the customer premises. DSL modem can terminate ADSL, SHDSL, VDSL circuits and provide LAN interface to end devices at private or commercial premises. Now-a-days, most DSL modems are integrated into the home routers. Thereby, the clear demarcation of local area network (home network) and the infrastructure of the Internet Service Provider (ISP) is no more possible.  DSL modem is connected to a digital subscriber line access multiplexer (DSLAM) port in telephone exchange using telephone company wires.  DSLAM converts analog electrical signals from subscriber local loop to digital data traffic (upstream traffic for data upload) and digital data traffic to analog electrical signals to subscriber local loop (downstream for data download). The traffic from DSLAM is routed to and from a broadband remote access server (BRAS) sitting at the edge of an ISP's core network, and aggregates user sessions from the access network. BRAS has following specific tasks
\begin{enumerate}
	\item Provides layer 2 connectivity through transparent bridging or Point to Point Protocol sessions over Ethernet (PPPoE) or ATM sessions. A BRAS can terminate upto 50000 PPPoE sessions.
	\item Inject policy management and Quality of Service (QoS).
	\item Provides layer 3 connectivity and routes IP traffic through the Internet Service Provider (ISP) backbone infrastructure to the Internet.
	\item Responsible for assigning network parameters such as IP addresses to the clients. BRAS is connected to following major necessary servers to assign network parameters to the DSL modem:
\begin{itemize}
	\item Authentication, authorization and accounting systems of ISP
	\item Auto Configuration Server (ACS) to perform remote management of the DSL modems at the customer premises.
	\item Domain Name System (DNS) server of ISP to translate the domain name address to IP address to enable the users to transfer data over the IP networks.
	\item Network Time Protocol (NTP) server to provide timing information to DSL modems.
	\item Dynamic Host Configuration Protocol (DHCP) server to lease unique public IP address to each DSL modems. 
\end{itemize}
\end{enumerate}
BRAS is technically the first IP hop from the client to the internet.  BRAS is connected to Internet router of the ISP to  route the user traffic into the Internet. Data traffic from one ISP is forwarded to the infrastructure of another ISP through their respective edge routers. The main functionality of Internet is to be used by users to transfer data from end devices over a medium (wired or wireless) through the infrastructure of an Internet Service Provider (ISP) to either other user's end devices or to servers of service providers. The end user will need the IP address of the destination to transfer the data through the Internet, this information can be obtained by making a domain name look-up using the services of DNS server of the ISP to obtain the IP address. The DNS server of ISP, if the requested domain name is not found locally, would recursively querying the authoritative DNS hierarchy. To illustrate the working of DNS, let's imagine a user wants to visit the website "www.amazon.com". The web browser first looks in its own cache for the IP address of the domain name "www.amazon.com", if no entry is found, the web browser will ask the operating system (OS). If OS doesn't have the entry, OS will perform the DNS query to DNS servers of ISP. The DNS servers of ISP are mostly configured to provide recursive DNS look-up. This means, the ISP's DNS server will quest the "root" name server to find out which name servers are authoritative (in simple terms responsible) for the ".com" domain. Once the information is available, ISP's DNS sever will send a query to those name servers asking them about the authoritative name server for "amazon.com" domain and then a final query to those name servers asking for the address of "www.amazon.com". The servers which answer for "www.amazon.com" and resolve the domain name to an IP address are called authoritative DNS servers. These authoritative DNS servers can be either company's private DNS server or a company hires a DNS service provider like Dyn to have managed DNS service. The figure \ref{fig:DNSOperations} illustrates the working of recursive DNS query processing without the consideration of DNS caching, web-browser caching or OS caching.

The another important component of Internet is Content Delivery Networks (CDN). CDN operators like Akamai host servers all over the globe and offer the service to cache bandwidth hungry contents for the websites for some nominal charge so that the end users can load those contents from one of their servers closest to users. Organizations depend on these CDNs to store and distribute contents to their customers around the world so that customer's servers will not be overwhelmed by user traffic and also to enable speedy delivery of bandwidth intensive contents to the end-users.

CDN, DNS Service providers and ISP are the key players who offer the services marketed as Anti-DDoS or DDoS mitigation services where the traffic are beneficially filtered or absorb and redistribute flood of malicious traffic. 

\subsection{Protocols}
\label{sec:tcp}

When the user data travels through the Internet, it propagates independent of the medium (e.g., copper wires, fiber optics cables, mobile networks, etc) because protocols are defined to separate from the means of communication. Protocols are high-level abstractions of network communications which enables to abstract details of the physical medium. Networks are built on layered communication architecture known as protocol stack. There two primary protocol stacks widely used in reality, namely, Open Systems Interconnection (OSI) and the Transmission Control Protocol and Internet Protocol (TCP/IP) architecture. OSI architecture has 7 layers and TCP/IP architecture has 4 layers. TCP/IP architecture can be seen as a simplified version of OSI architecture where layers 1,2 and  layers 5,6,7 of OSI is mapped directly to layer 1 and layer 4 of TCP/IP architecture, respectively. TCP/IP architecture is widely used in comparison to OSI architecture so we will restrict our discussion to TCP/IP architecture. The layers in TCP/IP architecture, their responsibilities and the most common protocols relevant for our discussion are listed in the table \ref{tab:tcp}.

\begin{figure}
\begin{center}
  \includegraphics[width=0.48\textwidth]{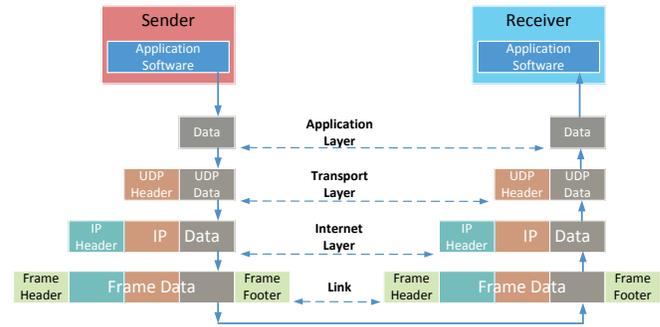}
\end{center}
\caption{Layered communication over TCP/IP protocol stack with UDP as transport layer protocol}
\label{fig:TCPLayer}       
\end{figure}

%
\begin{table*}
\caption{TCP/IP Protocol Layers}
\label{tab:tcp}       
\begin{tabular}{p{1cm}p{2cm}p{7cm}p{6.4cm}}
\hline\noalign{\smallskip}
Layers & Name & Responsibilities & Common Protocols  \\
\noalign{\smallskip}\hline\noalign{\smallskip}
4 & Application & User interaction, data presentation, & HTTP, SOAP, FTP, Telnet, DNS, SSL/TLS, SSH, DHCP, BGP, NTP, SNMP, TR-069 \\
3 & Transport & Convert messages to packets, sequencing, integrity, flow control, end-to-end error detection and correction & TCP, UDP \\
2 & Network & Convert packets to datagrams, routing & IP, ICMP, ARP, GRE (L3 tunneling protocol) \\
1 & Physical & Convert datagrams to bits, bit transmission & DSL, Ethernet, Optical fiber \\
\noalign{\smallskip}\hline
\end{tabular}
\end{table*}
Details about each protocols listed in the table \ref{tab:tcp} can be found in table \ref{app:protocols}. The figure \ref{fig:TCPLayer} illustrates the layered communication over the TCP/IP protocol stack between data sender and receiver. Layers in the protocol stack communicate in three directions. Each layer communicate abstractly with the layer above thereby abstracting the working of the current layer to the layer above. Similarly, the layer communicates less abstractly with the layer below and while the data is communicated to parallel or across the same layer in the protocol stack of the communicating peer. Interaction between the above and below layers are actual interaction between the layers, while the interaction between the parallel layers are virtual communication. Each layer adds its own activity to enable communication between two peers. The logical data communication path propagates from layer 4 to layer 1 for a sender and from layer 1 to layer 4 for a receiver. Physical communication always occurs at the layer 1. Peer-to peer correspondence occurs only at the like layers of sender and receiver. For example, if the sender's layer 3 affixes a header to the user data, then the receiver's layer 3 reads the header and removers it after verification that the receiver is the intended recipient.

\section{DDoS Attacks and IoT Botnets}
\label{sec:ddos}

\begin{figure}
\begin{center}
  \includegraphics[width=0.48\textwidth]{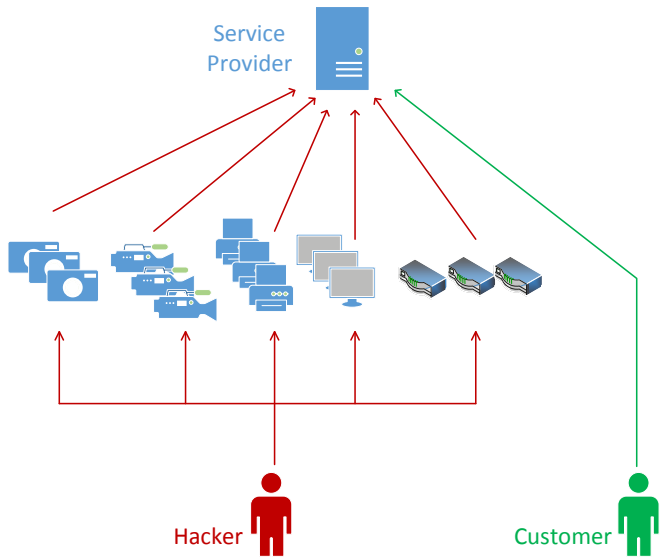}
\end{center}
\caption{Distributed Denial of Service (DDoS)}
\label{fig:DDoS}       
\end{figure}

Distributed Denial of Service (DDoS) attack is one of the most interesting and widely seen cyber attacks in the recent times. In DDoS attack, a hacker temporarily enslaves a number of internet-enabled devices into an arrangement known as botnet and then make simultaneous requests to a server or a array of servers for a specific service, thereby overwhelming the server and make it ignore legitimate requests from end-users (see Fig: \ref{fig:DDoS}). In simple terms, imagine a group of random people crowding up at the entrance or the shop show room space making no room for the geniune customers to enter the shop, thereby interrupting the business operations of the shop. A hacker can do this for different reasons, earlier it was for bragging rights, but now-a-days, these attacks are carried out by organized criminals for financial gains or for revenge or extortion or activism. 

A DDoS attack can exhaust the bandwidth (communication medium) or resources of the victim. Bandwidth can be expensive for small and medium enterprise. Most servers today are capable of serving 1Gbps, costs 50\$ per month on average or a 10Gbps server costs 300\$ per month on average. Exhaustion of the bandwidth for a period of time would lead to unavailability of the server during that time. Resources of server can be either processing capacity, number of ports, memory (RAM or persistent storage), processing capacity or others. Exhaustion of resources of the server can lead to either unavailability of the server or also lead to undesirable state of the server during which the cyber criminals can compromise the data in the server. DDoS attacks are often performed by two techniques, namely, reflection and amplification technique. In reflection technique, the attacker sends different packets with forged IP address of the victim / target as the source address of the packets to different destinations, which lead to the destination servers responding to the victim / target with the response to the sent packets from the attacker. Reflection technique is used by the attackers to usually hide his / her trail. In amplification technique, the small number of packets from attackers will elicit a large number of packets directed to the victim / target system. Amplification technique is often combined together with the reflection technique to lunch a large attack on to an unsuspecting victim / target. Malformed, unrequested, or recursive TCP, UDP, or ICMP traffic are the most common type of traffic used in a DDoS attack. In particular, for the reflection and amplification attacks DNS, NTP and SNMP traffic are favored\cite{5}. 

Another important aspect about DDoS attack is that the threat actors now-a-days deploy a layered attack with multiple attack vectors to achieve their goal. Therefore, the real intention for a DDoS attack is difficult to identify. Apart from disrupting the daily operations, DDoS can be used to probe the defenses of a victim / target or to just distract the victim / target during a actual attack using a different attack vector or exploiting a different vulnerabilities. Verisign in its report \cite{3} has found that, between 1st April and 30th June 2016, there has been 64\% DDoS attacks employing multiple attack types.

One can observe from the recent cyber incidents that the vulnerabilities of IoT devices are most effectively utilized by botnets to launch wide range of distributed denial of service (DDoS) attacks. In \cite{IoA}, researchers have found that most DDoS attacks in recent time originate from 3 types of devices, almost 96 percent were IoT devices, approximately 4 percent were home routers and less than 1 percent were compromised Linux servers. The IoT botnets not only affect the owners of the IoT devices but also anyone on the internet. To understand the IoT malwares, we need to understand the constrains posed by the environment of IoT devices, which have the following characteristics:
\begin{itemize}
	\item Embedded Linux dominate the IoT landscape with wide variety of libc implementations / version along with same ABI-compatible Linux kernel (2.4 < x < 4.3)
	\item Small memory capacity (RAM)
	\item Limited flash capacity mostly used to store embedded OS and firmware
	\item Non x86 architectures, mostly ARM, MIPS
	\item Support Executable and Linkable Format (ELF) binaries
	\item Rarely has any integrated User Interface (UI)
	\item Networked (mostly also internet enabled)
\end{itemize}

The threats to these IoT devices are important concern because they are hard to re-mediate and fix. IoT devices are low hanging fruits for the bad guys. The existence of IoT botnets has been a known fact since 2008. However, the extent of danger posed by the them was not realized until the second half of 2016. The key characteristics of IoT malwares used to orchestrate DDoS attacks are the following:
\begin{itemize}
	\item Most of the IoT malwares are Linux based malwares.
	\item Majority of the IoT malware has limited or no side-effects on performance of the host. They become active and perform DDoS on certain command from its botnet herders. 
	\item Many IoT malware reside on IoT devices' temporary memory (RAM). 
	\item Most IoT malwares does not use reflection or amplification techniques to launch an attack, so it is much difficult to recognize and mitigate the attack using the conventional methods.  
	\item Volume of traffic floods generated by IoT botnets are very high, in the orders of 100 Gbps or higher, in comparison to conventional PC botnets. 
	\item The location of the infected IoT devices are distributed all around the world (see figures \ref{fig:carna} and \ref{fig:mirai} ).
	\item Apart from generating commonly used traffic floods, namely,  HTTP , TCP, UDP traffic, some IoT botnets generates unconventional traffic like GRE traffic and use uncommon "DNS water torture" technique during DDoS attacks. 
\end{itemize}	
	
The use of GRE traffic flood is very unusual for DDoS attacks especially because source of GRE traffic cannot be forged. Most Internet routers allow GRE traffic as they are used to generate Virtual Private Networks (VPN) connections. GRE is also used by DDoS scrubbing providers as part of the mitigation architecture. 

\begin{figure}
\begin{center}
  \includegraphics[width=0.48\textwidth]{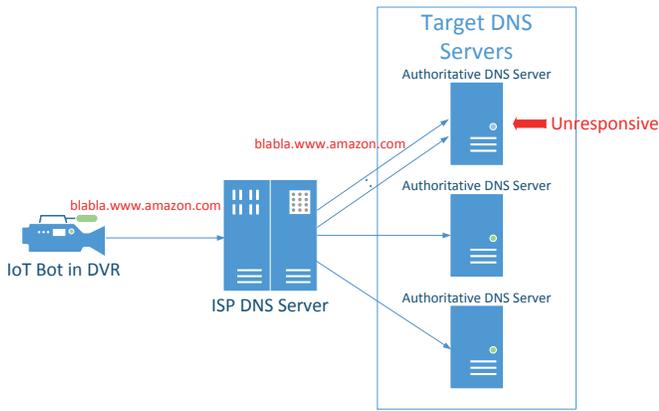}
\end{center}
\caption{DNS Water Torture Technique}
\label{fig:WaterTorture}       
\end{figure}

The uncommon "DNS water torture" technique illustrated in the figure \ref{fig:WaterTorture} is different from the conventional amplification or reflection techniques as it requires significantly less queries to be sent by the bot, letting the ISP's recursive DNS server perform the attack on the target's authoritative DNS servers. In this technique, a well-formed DNS query containing the target's domain name along with a randomly generated prefix appended to the name is sent by the bot. ISP's DNS server will recursively query the authoritative DNS server, if one gets overloaded, the request is forwarded to the next authoritative DNS server of the organization. Thereby, ISP's DNS server launching the attack on behalf of the IoT bot.

\subsection{Evolution of IoT Malware}
\label{sec:evo}

In this section, we will briefly outline the evolution of IoT malwares responsible for DDoS attacks in the recent times. The list of IoT botnet malwares discussed below is not complete especially because script kiddies and cyber criminals have been modifying and updating the known malwares to exploit new vulnerabilities or to infect diverse types of IoT devices. 

\begin{figure}
\begin{center}
  \includegraphics[width=0.48\textwidth]{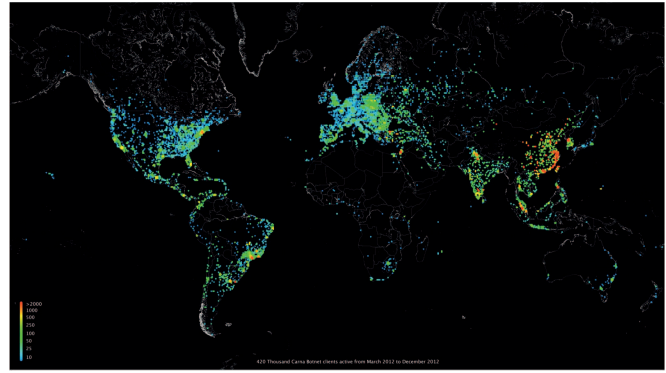}
\end{center}
\caption{Distribution of Carna Botnet between March 2012 to December 2012 (Source: \cite{carna})}
\label{fig:carna}       
\end{figure}

\begin{itemize}
	\item \emph{Linux/Hydra} is the earliest known malware targeting IoT devices. It is an open source botnet framework released in 2008. It was designed for extensibility and features both a spreading mechanism and DDoS functionality \cite{infodox,junus}. 
	\item \emph{Psyb0t} was found in-the-wild targeting routers and DSL modems in 2009. Estimated 100000 compromised devices were infected by the malware. Psyb0t is operated by Internet Relay Chat (IRC) based command-and-control servers. The primary methods to infect IoT devices used by Psyb0t are Telnet and SSH access using simple brute force attack with predefined 6000 usernames and 13000 passwords. \cite{psybot}
	\item \emph{Chuck Noris} is a IRC bot found infecting routers and DSL modems in 2010. Similar to Psyb0t, it was spreading by brute forcing passwords but also could exploit authentication bypass vulnerability in D-Link routers \cite{mcmillan}
	\item \emph{Tsunami} is another IRC bot which modifies the DNS server setting in the configuration of the infected devices such that the traffic from IoT device is redirected to malicious servers controlled by the attacker \cite{junus}
	\item \emph{LightAidra/Aidra} is a IRC-based mass scanning and exploitation tool support on several architectures, namely MIPS, MIPSEL, ARM, PPC, x86/86-64 and SuperH. Malware is designed to search open telnet ports that could be accessed using known default credentials \cite{fitsec}. The source code of LightAidra is freely available on the Internet as open source project \cite{AidraSource}
	\item \emph{Carna} is a botnet created by an anonymous hacker to measure the extent of the Internet and to get an estimation of the IP address usage. The data was collected by infecting Internet enabled IoT devices, especially routers with empty or default credentials \cite{carna}.  Carna usually scans for LightAidra on infected IoT devices and attempts to remove files and block any ports used  by LightAidra for communication. The distribution of Carna Botnet between March 2012 to December 2012 is shown in the figure \ref{fig:carna}
	\item \emph{Linux.Darlloz} is termed as IoT worm by Symantec \cite{hayashi1} that spreads by exploiting an old PHP vulnerability to access a system and privilege escalation through default and common credential lists. Like LightAidra, it also supports various architecture including x86, ARM, MIPS, MIPSEL, PPC architectures. After infecting the device, it drops the telnet traffic via iptables configuration and terminates the telnetd process to block users from accessing the infected device using Telnet. Symantec found that Darlloz has infected more than 31000 devices by February 2014 \cite{hayashi2}. A newer version of Darlloz uses infected devices to mine crypto-currencies (Mincoins and Dogecoins) \cite{hayashi2}. Like Carna  botnet, Darlloz also targets specifically LightAidra. It attempts to remove files and block any communication ports used by LightAidra.
	\item \emph{Linux.Wifatch} is an open-source malware which infects the IoT devices with weak or default credentials. Once infected, it removes other malwares and disables telnet access while logging the message "Telnet has been closed to avoid further infection of this device. Please disable telnet, change telnet passwords, and/or update the firmware." in the device logs. Wifatch uses peer-to-peer network to update the malware definition and deletes remnants of malware which remain in the IoT devices\cite{wifatch}.
	\item \emph{TheMoon} is a IoT worm discovered by Johannes Ullrich of SANS in February 2014. This malware specifically targets Linksys routers and exploits a command execution vulnerability while parsing $'ttcp\_ip'$ parameter value sent in a POST request. Command-and-Control servers (C2s) of the malware is capable of using SSL for end-to-end communication with their bots \cite{themoon}. 
	\item \emph{Spike / Dofloo} is a backdoor/DDoS malware discovered around mid-2014 found targeting 32-bit and 64-bit Windows, Linux based PCs as well as IoT devices based on MIPS and ARM architectures. It has been used in several attacks aimed at organizations in Asia and the United States. Akamai observed that one of the attacks to have peaked at 215Gbps \cite{spike}. This malware appears to be developed by China based group and can launch attacks with various payloads including SYN floods, UDP floods, DNS query floods, and GET floods against targeted organizations.
	\item \emph{BASHLITE / Lizkebab / Torlus / gafgyt} is one of the popular malware which infects Linux based IoT devices to launch DDoS attacks. It was reported that BASHLITE is responsible for enslaving over 1 million IoT devices, constituting mostly of Internet enabled cameras and DVRs\cite{33}. It is capable of launching attacks of up to 400 Gbps . Most BASHLITE attacks are simple UDP, TCP floods and HTTP attacks. BASHLITE infect a IoT device by brute-forcing its telnet access using known default credentials. One interesting aspect of BASHLITE is that malware payload deployed in IoT devices has the BASHLITE's C2s IP addresses hard-coded into it and are easier to monitor. Most of the infected devices are located in Taiwan, Brazil and Columbia. The source code of BASHLITE was partly leaked in early 2015 and has led to many variants. BASHLITE is considered the predecessor of Mirai and is in direct competition for vulnerable IoT real estate.
	\item \emph{KTN-RM / Remaiten} is a IoT Malware which combines the features of Tsunami and BASHLITE \cite{remaiten}. It infects Linux based IoT devices by brute forcing using frequently used default usernames and passwords combinations from a list.  C2s interacts with the bots using IRC communications by an actual IRC channel. Remaiten is more sophisticated than Tsunami and BASHLITE derivative malware. Remaiten can adapt itself based on the IoT device architecture and the type of attacks it wants to launch \cite{remaiten}.  
	\item \emph{Mirai} is one of the most predominant DDoS IoT botnet in recent times. Mirai means "the future" in Japanese. Mirai botnet is definitely the next step in IoT DDoS botnet malwares, however not as sophisticated as Remaiten but most effective. At its peak, Mirai infected 4000 IoT devices per hour and currently it is estimated to have little more than half a million infected active IoT devices. Mirai botnet is famous for being used in the record breaking 1.1Tbps DDoS attack with 148000 IoT devices. Mirai targets mostly CCTV cameras, DVRs, and home routers. The source code for Mirai has been published by its alleged author Paras Jha \cite{mirai} using his online pseudonym "Anna-senpai" on the English-language hacking community Hackforums as open-source \cite{MiraiSource}. Since the release of the Mirai source code, the number of IoT infected devices has increased from 213000 to 483000 in just two weeks. Based on the IP addresses one can identify that the Mirai infected IoT devices are distributed in over 164 countries with highest densities in Vietnam, Brazil, US, China and Mexico (see figure \ref{fig:mirai}). The strength of the DDoS attack ranges from 200 Gbps to 1.2 Tbps. Mirai generates floods of GRE IP, GRE ETH, SYN and ACK, STOMP, DNS, UDP, or HTTP traffic against a target during a DDoS attack \cite{34,35}. More recently \cite{WindowsMirai}, Mirai has been found to be enhanced to infect Windows devices, helping hackers hijack even more devices. This enhanced Mirai malware could also identify and compromise database services like MySQL and Microsoft SQL running on different ports to create new admin "phpminds" with the password "phpgodwith" allowing the hackers to steal the database. The awareness of IoT botnets in recent times attributes to Mirai and the volume of traffic generated during its DDoS attacks. 	
	\item \emph{Linux/IRCTelnet} is a new IRC botnet ELF malware aimed at IoT devices with IPv6 capabilities. IRCTelnet combining the concept of Tsunami for IRC protocol, BASHLITE for the infection techniques (telnet brute force access and code injection) and using the Mirai botnet's IoT credential list. The base source code of LightAidra/Aidra is used to build the new botnet malware. The botnet is using UDP, TCP flood along with other series of attack methods in both IPv4 and IPv6 protocol. The new malware features the extra IP spoof option in both IPv4 or IPv6. 	
\end{itemize}

\begin{figure}
\begin{center}
  \includegraphics[width=0.48\textwidth]{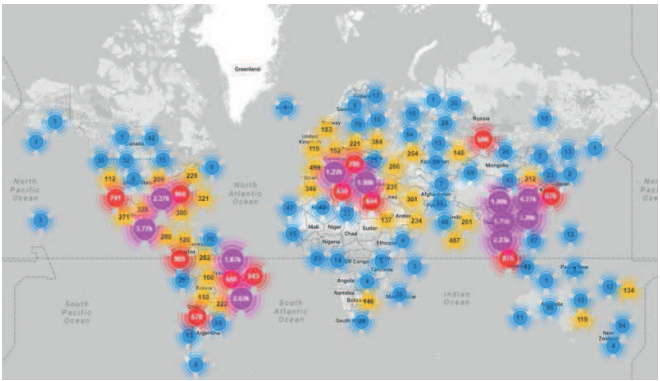}
\end{center}
\caption{Distribution of Mirai Botnet until 26th October 2016 (Source: \cite{35})}
\label{fig:mirai}       
\end{figure}

\subsection{DDoS-as-a-Service}
\label{sec:service}

In their report "State of Internet" (Q3 2016) \cite{akamai}, Akamai has reported to have observed 138\% increase (in comparison to 2015) in DDoS attacks with traffic greater than 100 Gbps attributing to the increase in the usage of IoT devices in DDoS attacks and the easy availability of commercial DDoS services or DDoS-as-a-Service (DDoSaaS) for hire. One can anonymously order a 5-6 Gbps DDoS attack lasting 10 minutes or more for as little as \$6. The DDoSaaS (also known as "booter"' or "stresser") providers sell the attack capabilities for knocking websites offline or perform stress tests on different network infrastructures. Recent study \cite{DaaS} found over 435 booter and stresser websites on the open Internet or Clearnet. However, there are much more offers on the Darknet. Darknet is an overlay network accessible using specific software, configurations and authorization which often uses non-standard communication protocols and ports. Darknet is often used by cyber criminals to offer their services. Some of the marketplaces available on Darknet are HELL, Alphabay, Valhalla, Hansa and The Real Deal among others. There marketplaces provide one click shopping experience for DDoSaaS, tools and personally identifiable information (PII) lists. The cost of a DDoSaaS varies based on the size of the botnet, type of attack, the victim or target's defenses, exclusivity of the malware and other factors. Two of the most famous DDoSaaS are listed below:

\begin{itemize}
	\item Shenron Attack Tool - provided by popular hacker group Lizard Squad as public stresser services. Eight different packages are available for interested customers, the cheapest DDoS package is offered for \$ 19.99 to launch a 35 Gbps attack for 20 minutes with UDP and TCP traffic. 
	\item vDOS Attack Tool - provided by hacker group, vDOS. It is one of the most preferred tools. Thirteen different attack vectors available for DDoS Attack.  Four different packages are available for interested customers, the cheapest DDoS package is offered for \$19.99 to gain access to 216 Gbps attack shared network.
\end{itemize}
In \cite{akamai} Akamai observe that the DDoSaaS traffic continue to account for a large portion of the attack traffic in major attacks. In the next section, we describe some of the major and minor IoT botnet (mostly Mirai) DDoS incidents in recent times to indicate the seriousness of the current situation.

\subsection{Recent Famous DDoS Incidents by IoT Botnets }
\label{sec:incidents}

\subsubsection{KrebsOnSecurity.com}
\label{sec:krebs}
On the evening of 30th September 2016, the blog of the security researcher, Brain Krebs, experienced a 623Gbps DDoS attack from a large number of compromised IoT devices\cite{6}. This particular DDoS attack was considered to be a retaliation action from a group of hackers due to Krebs' series of articles on the take down of the Israeli DDoS-as-a-Service provider called vDOS, which coincided with the arrests of two 18 years old men named in those articles as the founders of the service \cite{7}. Krebs was a pro bono customer of Akamai, a leading CDN operator which hosted his blog website and also provided the DDoS attack protection. After resisting the attack for three days as the DDoS mitigation efforts were beginning to cause problems to Akamai's paying customers, Akamai terminated their pro bono contract with Krebs. The DDoS attack was so aggressive that the Akamai platform couldn't handle the resources needed to mitigate the attack as it is nearly twice the size of the next-largest attack they had ever seen before \cite{8}. Had Akamai continued their free support, they will have to spend millions of dollars in cyber-security services. Akamai observed in their report \cite{akamai} that they have seen 5 different DDoS attacks on the blog in September 2016 alone ranging from 123 Gbps to 623 Gbps and see these DDoS attacks as the means used by cyber criminals to silence their detractors. After the abrupt termination of pro bono agreement with Akamai, traffic destined for the blog was redirected to 127.0.0.1 — effectively relegating all traffic destined for KrebsOnSecurity.com into a giant black hole. The blog is back online with the support of Project Shield, a free service run by Google to help protect news sites and free expression from online censorship through DDoS attacks.

According to the Akamai's Q3 2016 State of the Internet report \cite{akamai}, the DDoS attack was launched by just 24000 IoT devices from around the world, mostly constituting digital video recorders (DVR) and IP cameras that are exposed to Internet. The 623 Gbps DDoS attack was launched with the help of IoT devices infected with Mirai and BASHLITE malware \cite{6}. The attack traffic was mostly of GRE traffic and junk web traffic such as SYN, GET and POST floods from legitimate connections between attacking host and target.

\subsubsection{OVH}
\label{sec:ovh}

OVH is french cloud computing and hosting company that offers virtual private servers (VPS), dedicated servers and other web services. On 22rd September 2016, OVH founder and CTO Octave Klaba posted the message along with a screen-shot on his twitter account explaining that OVH servers are experiencing a series of DDoS attacks and many of them larger than 100 Gbps, with the severest single attack documented by OVH reached 799 Gbps \cite{9}. Later on 23rd September 2016, Klaba posted again on his twitter account that OVH has experienced a record breaking DDoS attack generating at least 1.1 Tbps to 1.5 Tbps traffic from 145607 camera / DVR with each IoT device sending traffic between 1-30 Mbps. The attack is suspected to be from IoT devices infected with Mirai and BASHLITE malware sending traffic mostly of type TCP/Ack, TCP/Ack+PSH and TCP/SYN \cite{9}. There was no reason given for OVH attack, however it is the largest DDoS attack launched so far. The hacker with pseudonym "Anna-senpai", who released Mirai botnet malware source code claims to have lived in France during the OVH attack and was avoiding law enforcement authorities for his involvement in DDoS attack on KrebsOnSecurity.com.

\begin{figure}
\begin{center}
  \includegraphics[width=0.48\textwidth]{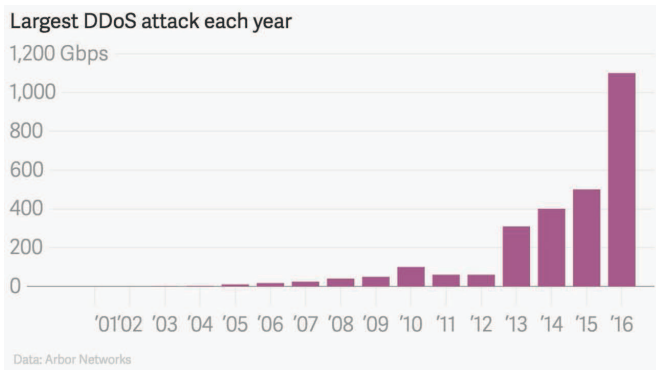}
\end{center}
\caption{Largest DDoS attack each year}
\label{fig:DDoSIoT}       
\end{figure}

\subsubsection{Dyn}
\label{sec:dyn}

Dyn is an Internet performance management company offering wide range of products to monitor, control and optimize online infrastructure. Dyn is also one of the leading managed DNS provider in the world. On 21st October 2016, Dyn experienced DDoS attacks on their managed DNS servers from 100000 Internet enabled IoT devices such as printers, IP cameras, residential gateways and baby monitors that generated masked TCP and UDP traffic over port 53. IoT devices used in the attack are found to be primarily infected by Mirai botnet malware \cite{10}. According to some experts, the attack magnitude reached in the 1.2 Tbps range but there has been no confirmation from Dyn \cite{10}. The attack has been analyzed as a complex and sophisticated attack that generated, apart from the masked TCP and UDP traffic floods, also compounding recursive DNS retry traffic which further amplified the impact of the attack. i.e., after the attack, Dyn DNS servers experienced further 10-20 times the typical amount of legitimate traffic from millions of IP addresses because the multiple DNS requests has been generated by recursive DNS servers from the user retries to access websites. The DDoS attack itself was launched at two  different time intervals, first attack begun at 11:10 UTC to 13:20 UTC  and then again from 15:50 UTC to 17:00 UTC. The first attack began at 11:10 UTC at the managed DNS platform of Dyn in Asia Pacific, South America, Eastern Europe and US-West regions which triggered a response from Dyn to initiate their incident response protocols. Dyn observed that the attack abruptly changed its target to their point of presence (POP) in the US-East region. Dyn's engineering and network operations team started to deploy additional mitigation mechanisms like traffic shaping, re-balancing of incoming traffic by tuning the anycast policies, internal traffic filtering and deployment of scrubbing services. The attack subsided by 13:20 UTC either because of Dyn's mitigation efforts or the attacker intended it so is not clear. A second wave of attack on Managed DNS platform started again at 15:50 UTC from more globally distributed IoT devices, however the mitigation mechanism already set in place due to first wave of attacks could be extended for the second attack and Dyn could substantially recover by 17:00 UTC. The residual impact from additional sources was observed until 20:30 UTC. Dyn reported to have observed a number of probing smaller TCP attacks during the next several hours and days after the two waves of DDoS attack.

\begin{figure}
\begin{center}
  \includegraphics[width=0.48\textwidth]{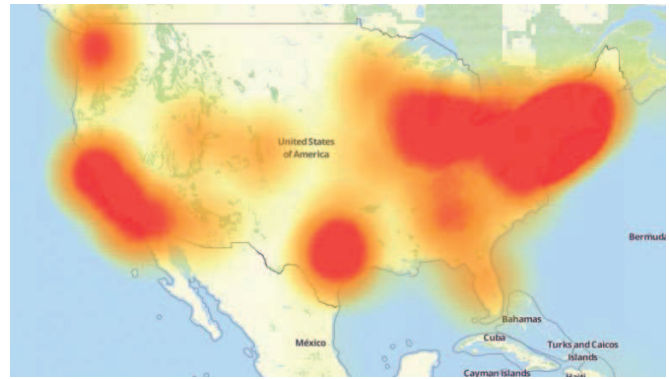}
\end{center}
\caption{Depiction of the outage caused in US by Dyn DDoS attack}
\label{fig:DynDNSImpact}       
\end{figure}

Dyn provide their Managed DNS service to some of the famous websites like, Airbnb, Amazon.com. Reddit, Spotify and many others who were partly or completely unavailable to large swathes of users in Europe and North America. The impact of the outage due to DDoS attack on Dyn is illustrated in the figure \ref{fig:DynDNSImpact}. According to Datanyze, leading techno-graphics provider which does analysis of Managed DNS market share, Dyn was the most favored Managed DNS provider offering their service 137 out 1000 top Alexa websites in 2015 but after the DDoS attack they now offer their service to only 90 out 1000 top Alexa websites and lost many customers to their main competitor Cloudflare DNS and Amazon Route 53 who market their service as Managed Cloud DNS service. 

The motivation behind the Dyn DDoS attack is not very clear. However, Brain Krebs notes in his blog \cite{11} that just hours before the Dyn attack, Doug Madory, a researcher working for Dyn, made a presentation to North American Network Operators Group (NANOG), based on a joint research with Krebs, about a dubious DDoS mitigation company called BackConnect. Madory indicated that BackConnect may have contacts with cybercriminal and uses non-ethical methods like BGP hijack to perform DDoS mitigation. There is no conclusive evidence for the claim of Krebs. Regardless of the reason, the Dyn incident made people to take notice of IoT botnets and DDoS attacks. It also inspired many discussion in cybersecurity and legislative communities. 

\subsubsection{Deutsche Telekom}
\label{sec:dt}
Deutsche Telekom is a german telecommunication company having strong presence in Europe and US. They are one of the most favored Internet service provider in Germany. Deutsche Telekom provide their DSL subscribers a home routers with in-built DSL modem, manufactured by different companies, under the brand name "Speedport". On Sunday, 27th November 2016, a large number of Deutsche Telekom customers reported connectivity problems. These issues were traced to particular models of "Speedport" home routers, namely, Speedport W 921V, Speedport W 723V Typ B, Speedport W 921 Fiber manufactured by Taiwanese manufacturer Arcadyan. The root cause of the problem was found to be the new Mirai malware variant which was trying to actively scan and infect the vulnerable devices to expand the number of infected devices in their botnet. The vulnerability intended to be exploited by these new Mirai malware variant was reported by a security researcher "kenzo2007" on his blog dated 7th November 2016. In his blog entry \cite{eir}, "kenzo2017" showed that TR069/TR064 was not properly implemented in D1000 modem manufactured by Zyxel and used by Irish ISP Eir. Interestingly, Taiwanese manufacturer Arcadyan who produced vulnerable "Speedport" home router for Deutsche Telekom, appears to be not connected to Zyxel, the makers of the vulnerable Eir modem. 

There are three different vulnerabilities exploited by the new Mirai malware variants.
\begin{enumerate}
	\item Most ISP leave the port 7547 open on ISP supplied home router / modem for remote management of CPE using TR069 (CPE WAN Management Protocol). Though, this is not a huge flaw, it is not recommended as anyone can access CPE (router/modem) using the port 7547. The authentication method used by TR069 either requires no passwords or use weak HTTP digest authentication method over unencrypted path or using certificate authentication, which is in most cases not implemented properly by the manufacturers. This issue was first pointed out by security researcher Luka Perkov in his talk "ISP Black Box" at 28C3 and more recently by Shahar Tal in his talk " I Hunt TR-069 Admins: Pwning ISPs Like a Boss" at DEFCON 22. A simple search on Shodan (www.shodan.io) shows approximately 41 million devices have their port 7547 open.
	\item TR064 ( LAN-Side DSL CPE Configuration) Server was running behind the port 7547 meant for TR069. TR064 strictly meant for local configuration of CPE but not for remote management. This flaw allows anyone on Internet to access CPE (Router/Modems) and perform important device (CPE) configurations.
	\item The crucial flaw in these specific models of CPE is that "SetNTPServer" command of TR064 can be used to execute arbitrary commands (command injection vulnerability).
\end{enumerate}

\begin{figure}
\begin{center}
  \includegraphics[width=0.48\textwidth]{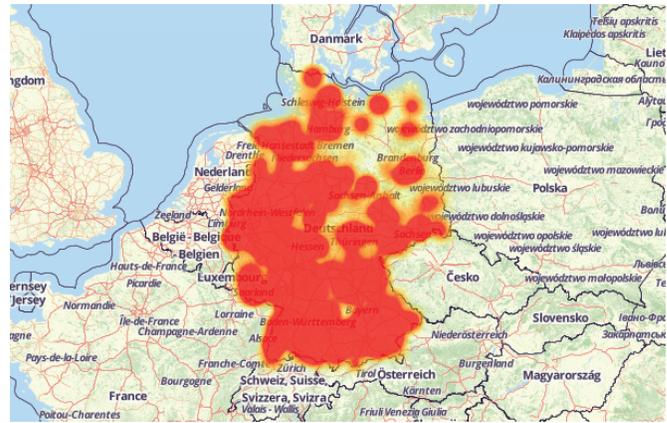}
\end{center}
\caption{Depiction of the outage caused in Germany by Mirai attack on Deutsche Telekom Customers}
\label{fig:DTMirai}       
\end{figure}

In \cite{comsecuris} Comsecuris, a security company, ran some tests on one of the "Speedport" modems and found it is not vulnerable, but they did observe that the modem became slow or non-functional even under moderate load, so it is possible that even-though the new Mirai malware variant was not successful, it caused the modem to crash. Mira IoT malware resides on temporary memory (RAM) of IoT devices, therefore a simple power reset can remove the malware from the infected IoT devices. However, many IoT botnet herders constantly scan the internet for vulnerable Internet enabled IoT devices. As per SANS, a non-profit security institution, the new Mirai variant was continuously scanning the Internet for new vulnerable devices and could find any newly connected router/modem within 10 minutes. Tests done by security researcher Darren Martyn show that modems used by UK ISP TalkTalk, D-Link DSL 3780 modems, modems made by MitraStar, Digicom and Aztech are all vulnerable. He states that he found 48 different vulnerable devices in use now. Deutsche Telekom informs that the incident affected 900000 home routers, which is 1 in 20 customers of Deutsche Telekom in Germany. Similar incident has been reported by UK ISPs TalkTalk and UK Post Office's Internet broadband service affecting over 100,000 customers. Deutsche Telekom and other UK ISPs were able to fix the security problems in vulnerable devices by providing firmware update. The impact of the outage in Germany due to the attack by new Mirai variant on vulnerable "Speedport" home routers is illustrated in the figure \ref{fig:DTMirai}.

\subsection{Other Notable Minor IoT Incidents}
\label{sec:minor}
Since the beginning of 2013, there has been many minor DDoS attacks by different IoT botnets. In the wake of the DDoS attack on Dyn, MalwareTech.com has set up a automated twitter account (@MiraiAttacks) that live tweets the information from their honeypot system, which includes the information about the botnet used, type of traffic, duration of the attack, target IP and port addresses. At the time of this article, there are 79 different botnets being monitored by the account \cite{12}. However, in this section we discuss some of the most interesting minor incidents happened in recent times. 

\subsubsection{CCTV vs Small Business Websites}
\label{cctv}
Sometime in June 2016, web security firm, Sucuri was contacted by a small jewelery shop to help them mitigate a DDoS attack their website was experiencing. They observed the website was under DDoS attack peaked at nearly 50000 HTTP requests per second and lasted for days. They could mitigate the attack swiftly by moving their DNS to Sucuri servers and soon realized that jewelery shop website was not the only one under attack, but thousands of websites were under DDoS attack by 25000+ compromised CCTV devices located all around the world. Researchers has observed 25000 unique IP addresses from locations like Taiwan, US, Indonesia, Mexico and Malaysia. The most interesting aspect of this attack is that 5\% of the IPs observed are IPv6 \cite{cctv}.

\subsubsection{Liberia}
\label{sec:liberia}
In early November, Mirai IoT botnet attack on telecommunication infrastructure of Liberia was observed by @MiraiAttacks twitter account. This led to the speculation that Internet in Liberia was brought down by Mirai \cite{13}. The Internet undersea cable to Liberia installed in 2011 was claimed by Beaumont as single point of failure for Internet access in Liberia \cite{14}. The speculation of the possibility of Mirai botnet knocking the Internet access to Liberia was further fueled by the claim of a security researcher reported from anonymous sources that an attack of 500 Gbps had targeted Liberia's undersea Internet cable. This coincided with network connectivity problems observed by the users in Liberia. The general manager of the Cable Consortium of Liberia reported that the African-Coast-to-Europe (ACE) submarine cable monitoring system and the servers locally located in Liberia Internet Exchange Point (LIXP) show no downtime during the time attack was reported. However, Kpetermeni Siakor, who manages the infrastructure at the LIXP, informed that Lonestar Cell MTN, one of the four major telecommunication companies, faced 500 Gbps DDoS attack for a short period but was successfully mitigated.

\subsubsection{Lappeenranta, Finland}
\label{sec:finland}
Two housing blocks in Lappeenranta, Finland experienced disruption of heating distribution between late October to 3rd November 2016. This incident is of particular interest as the disruption was attributed to DDoS attack by Mirai botnet \cite{17,18}. Though media outlets find the story of Finlands residents stranded in the cold of winter due to a malicious DDoS attack by Mirai on the remote systems, manufactured by Fidelex, controlling the central heating and hot water distribution is an enticing story but the truth may be different. This incident is similar to previously mentioned Deutsche Telekom incident, however the vulnerability exploited here is different and simple. The security of these building automation systems were ignored or neglected and were not behind any network firewall or other perimeter security. Fidelex confirmed that the vulnerabilities in the system are opened up when the operators configured the devices for convenience \cite{17}. The new IoT botnet malwares actively scanning the Internet for vulnerable IoT devices could easily reach them. In an attempt to mitigate the attack, the system automatically reboot the main control and got caught in an infinite restart loop that eventually led to the heating system being offline for more than a week \cite{17,18}. According to Valtia, a Lappeenranta based facilities management company, managing these remote heating system observe that over 90 percent of these remote systems in terraced house or large building in the area were built to be fail-safe, meaning if an error occurs these systems would shutdown the heating system and attempt to automatically reboot. The remote systems will also not send alarm if the heating system is switched off or radiator pressure just disappear.  Valtia finally identified the malfunctioning systems and switched the heating system to manual control, installed firewall to limit and filter the traffic and bought the control systems back online\cite{18}. Though there is no evidence for the involvement of Mirai botnet malware in this incident, the threat posed by the IoT botnet malware to critical infrastructure is very real and can physically affect people.

\subsubsection{Russian Banks}
\label{sec:russian}
In October 2015, eight different Russian financial institutions were targeted in a DDoS campaign to inflict significant impact on Russian economy. More recently, at least five Russian banks including Sberbank, Alfa bank, the Moscow Exchange, the Bank of Moscow and Rosbank experienced prolong DDoS attack between 8th November to 10th November 2016 from 24000 IoT devices distributed around 30 countries \cite{24,25,26}. Though the banks never public acknowledged the impact of these attacks, one journalist who actively monitors their websites could not reach their websites during the attack duration \cite{24,25}. A DDoS-as-a-Service (DDoSaaS) operator "vimproducts", claims to have conducted those attacks for his/her clients in retaliation for Russia's alleged involvement in US elections. "vimproducts" offers his/her DDoSaaS offer in different qualities on Alphabay to interested buyers.

\subsubsection{US Elections}
\label{sec:elections}
Flashpoint, Internet security company, reported three different incidents DDoS attack on campaign website of Donald Trump at 16:20 UTC on 6th November 2016, 8:13 UTC and 8:19 UTC on 7th November 2016. In the last instance, the attack was also targeted the site of Hillary Clinton. However, no outages were reported for either website. In all the three instances a 30 second HTTP application layer attack was launched using Mirai botnet. Flashpoint believe that each of these attacks may have been carried out by two different groups \cite{20}.

Another DDoS incident related to US elections reported to have occurred on 6th November 2016, targeted a phone bank service, TCN, used by election campaigns. The attack begun with with a small flood of junk traffic from a small pool of IP addresses, but soon progressively increased until all 4 of 1 Gbps connections of TCN got saturated. The attacker continued the DDoS attack for 24 hours, varying the IP sources and type of traffic flood generated \cite{21}. A 4chan user using the nickname "Sparky", claimed to have launched these DDoS attacks against the Clinton campaign phone lines. However, these attacks influence both Republican as well as Democrats campaigns. The attack overwhelmed TCN's infrastructure and periodically took their web-based software offline. TCN responded by procuring anti-DDoS protection from CloudFlare, which filtered the attack traffic and deploying a number of proxy servers designed to absorb excess traffic \cite{21}. 

\subsubsection{WikiLeaks}
\label{sec:wikileaks}
On 7th November 2016, the email publication servers of Wikileaks were knocked offline by DDoS attack lasting for nearly 24 hours. The attack was allegedly in response to the "DNSLeak2" where 8263 emails from the compromised account of John Podesta, chairman of the Clinton campaign was released \cite{22,23}. Regardless of the political motivations or correctness of the actions of Wikileaks, this incident shows that DDoS attacks are now-a-days often used as means to restrict the freedom of speech on the Internet. Google has started an initiative called Project Shield which provide free anti-DDoD services to websites that have "media, elections, and human rights related content". KrebsonSecurity.com is also one of the beneficiaries of this initiative. 

\subsubsection{Random Denial of Server (RDoS)}
\label{sec:RDoS}
In an RDoS attack, the cyber criminals send a letter or email threatening to attack an organization either by causing interruption to business, operation or cyber presence, unless a ransom is paid with a deadline. In a recent study \cite{rdos}, one in seven say they have has a ransom attacks in the past year. There are many Armada Collective and Lizard Squad impersonators send ransom emails and made use of DDoSaaS to send out some warning shots to make their demands fulfilled favorably. These impersonators have reportedly earned \$100000 with the wave of ransom letters \cite{rdos}.

\subsection{Anatomy of IoT Botnets}
\label{sec:anatomy}

\begin{figure}
\begin{center}
  \includegraphics[width=0.48\textwidth]{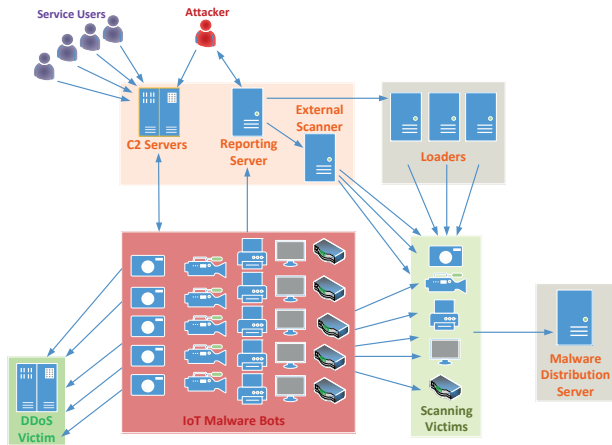}
\end{center}
\caption{Generic structure of an IoT botnet}
\label{fig:MalwareStructure}       
\end{figure}

In this section we will analyze the structure of IoT botnets and the mode of operations based on the publicly released source code of IoT botnet malwares, namely, LightAidra/Aidra, BASHLITE and Mirai as well as based on the information from the collective reverse engineering efforts made by many security researchers \cite{35,39,40,IoA}. The code of IoT malware are mostly written in C language. For example, BASHLITE has been written in C language. Mirai botnet uses both C and Google Go language. LightAidra/Aidra uses Python along with the C language. The source code for the bots is cross-compiled for multiple architectures running the Linux operating system which is common operating system among most of the IoT devices and embedded systems. The generic structure of the IoT botnet is illustrated in the figure \ref{fig:MalwareStructure}.

The IoT botnets consists mostly of the two basic and four additional components, namely,
\begin{enumerate}
   \item \textbf{\emph{Bots}} or agents or the end zombie IoT devices that perform DDoS attacks on command
   \item \textbf{\emph{Command-and-control servers (C2s)}} are used to control the bots
   \item \emph{Scanners} are used to scan for vulnerable IoT devices.
   \item \emph{Reporting server} used to collect the results or scan reports from bots or external scanner
   \item \emph{Loaders} used to log on to vulnerable IoT devices and instruct them to download malware
   \item \emph{Malware distribution server} is the location where malware code is stored to be downloaded by infected IoT devices
\end{enumerate}
The functions of one or the other components listed above can be combined and its function can be performed by another component. For example, in Mirai, there exists no separate scanner component, however the bots perform the function of scanning for vulnerable IoT devices and also carry out DDoS attacks on target. In a general scenario, C2s communicate regularly with bots, foot solders in botnet. Most botnets implements a standard client/server architecture where the bots get their commands from the C2s or controllers. The botnet malware spread to new IoT devices by continuously scanning the internet for vulnerable IoT devices, either from the bots or from an external scanners (in some cases, C2s performs the scan directly). Potential victims devices can be found using special search engines such as Shodan (www.shodan.io) and Censys (www.censys.io). Reporting server receives one-way traffic with information about the IP addresses and credentials of the vulnerable IoT devices from scanners (as in BASHLITE) or from the bots (as in Mirai malware). 

Most IoT devices have telnet and web interface enabled with default credentials for allowing the end users to access these devices. In most cases, the end users don't change the default credentials of the IoT devices. Unfortunately, these services with default passwords are the first point of access for IoT malwares. Loaders uses the information sent to reporting server by scanners to log on to the vulnerable IoT devices with the reported credentials. Most of the IoT malware uses the telnet connection to communicate with the vulnerable IoT devices. Once the loaders accesses the IoT devices, it instructs the devices to download the malware from a malware distribution server. Most of them use the command "wget" to download the malware binary. In the case of Remaiten, loaders can analyze the hardware architecture of the IoT device and download the necessary malware components to infect the device. After downloading, the execution of the malware binary enables the IoT devices to come under the control of the botnet herder. The botnet herder interacts with the C2s and reporting server for the duration of botnet lifetime. After the establishment of botnet, it may be sold to various users who may use an API hosted on C2s to perform DDoS attack. The botnet author may use TOR (anonymity network) to hide his source IP address. The some cases as in BASHLITE, malware in the bot may have hard-coded IP addresses of the C2s. In the case of Mirai, bot malware uses domain name to resolve the IP address of the C2s on the fly, that way IP address of C2s can be changed regularly to evade detection.

To summarize, the common mode of operation for most IoT botnets are as follows:
\begin{enumerate}
	\item Malware (external scanner or bots or C2s) continuously scans the internet for vulnerable IoT devices, usually for open Telnet ports or other open services reachable over Internet
	\item Once a vulnerable IoT device is found, malware (external scanner or bots or C2s) access the vulnerable IoT devices using brute forcing with list of known default credentials
	\item The device IP address along with the successful password are stored in a reporting server to be used by malware (loaders or C2s) to access the device later
	\item Malware (loaders or C2s) accesses the scanned IoT device with the credentials stored in reporting server. It exploits known security weaknesses in available services on the IoT devices to download additional malware payload from a malware distribution server.
	\item Bot malware becomes active by the execution of the downloaded malware binary. It escalates privilege by exploiting the known security weaknesses in IoT devices
	\item Bot malware secures the IoT devices by fixing the vulnerabilities used to access the device so that no other malware can access the same IoT device
	\item Competing malware, if found in the IoT devices, are eradicated by bot malware using different techniques like memory scraping among others
	\item Bot malware reconfigures the device to be part of the botnet 
	\item Bots communicate regularly with their C2s using IRC based protocol to indicate their existence to the botnet operator
	\item Malware in bots remain dormant and doesn’t noticeably affect IoT device performance or functioning, until a user or botnet owner instructs them perform a DDoS attack
\end{enumerate}

IoT malwares are becoming increasingly adaptive and sophisticated with many new features like IPv6 support, sophisticated communication methods between bots and C2s. The IoT landscape is also growing large with the decision of many consumer electronic manufacturers to produce more Internet enabled consumer electronics. The fight for dominance amongst different IoT malwares for the IoT real estate is a commonly observed characteristics of IoT malwares, for example, Carna, Darlloz and Wifatch removed LightAidra, Mirai removed BASHLITE and little known Anime malware. This aggressive behavior is to maximize the attack potential of the botnet devices and to prevent similar removal attempts from other malwares.

\section{Remediations and Recommendations}
\label{sec:RR}
\begin{figure}
\begin{center}
  \includegraphics[width=0.48\textwidth]{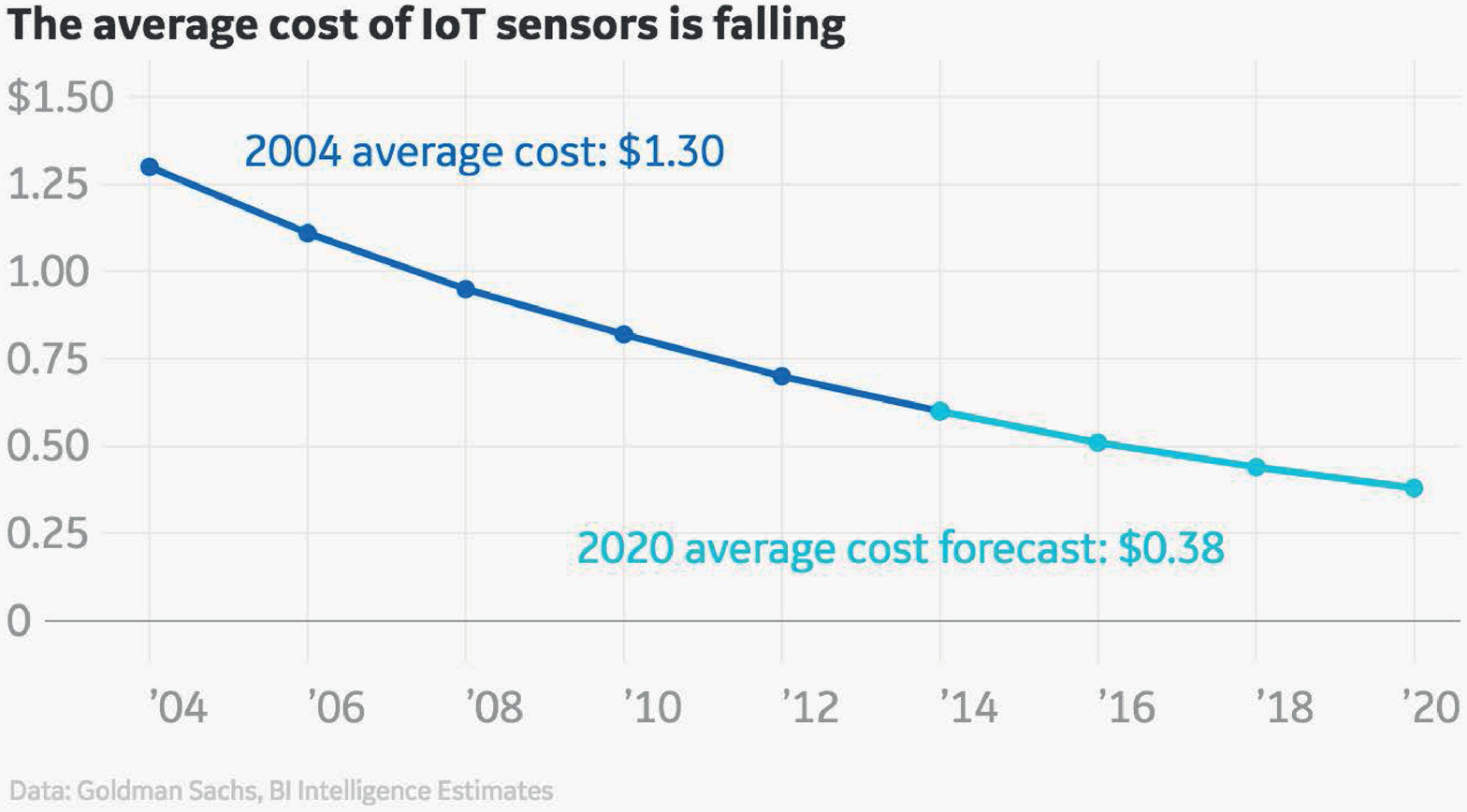}
\end{center}
\caption{The average cost of IoT sensors}
\label{fig:CostIoT}       
\end{figure}

In a recent studies \cite{HP}, HP found 70\% of the IoT devices to be vulnerable. The root cause for the vulnerabilities in IoT devices is the rush to bring new products and services to market by third party vendors. Moreover, these devices are produced in offshore with very low price margin (see Figure \ref{fig:CostIoT}), forcing manufacturers to concentrate on the functionality than the necessary security.  Many IoT device manufacturers ignore providing security updates during the device life cycle.  Most of the devices have limited or no means to be patched to fix known security bugs. Some of them even have hard-corded default user credentials, like in the case of DVRs and IP cameras made by a Chinese hi-tech company called "XiongMai Technologies" \cite{hardcoded}. The vulnerability in these devices will exist until they are thrown away and that will take a while. The life time for a IoT device varies but on an average longer than five years. For example, on an average a smart phone is replaced every 18 month, DVR lasts for 5 years, a car for 10 years, a refrigerator for 25 years, a thermostat will almost never be replaced in one's lifetime.  Therefore the security issues seen in existing IoT devices will not disappear in near future. The improvements in IoT security will have happen only when the IoT device manufacturers and end-users share responsibility of reducing risks. 

Some ideas to mandate IoT devices manufacturers and vendors to comply with several security measures like the following:
\begin{itemize}
	\item Hard-code the devices to limit communication to private IPv4 address (RFC 1918) or website of the manufacturer. Communication with all other IP or domains should be limited.
	\item Each IoT device should have an unique default password comprising of 10 or more characters with capital or small alphabets, numbers and special characters.
	\item By default, Internet enabled IoT devices should be required to connect to manufacturer's or vendor's website periodically to check for security updates. The functionality of the IoT device shall be reduced to bare minimum if the IoT devices doesn’t connect to the manufacturer's website for a given period of time. 
	\item End-user should activate the device with the manufacturer's or vendor's website by providing necessary user information like contact name, email address and other contact information. If the device doesn't contact the manufacturer's or vendor's website, they can inform the end-user using the provided contact information.
	\item Laws should be put in place to ensure that the manufacturers be held accountable for following and implementing security best practices in their devices. 
	\item End-users can protect themselves from participating in a DDoS attack through the procession of infected IoT devices by following some best practices (listed in Section \ref{sec:best}).
	\item Commercial companies who want to protect themselves from being a victim of DDoS attacks can buy DDoS mitigation services offered by ISPs and CDNs like Dyn, Akamai and others.
	\item Apart from implementing the best practices listed in Section \ref{sec:best}, commercial companies should develop actionable and practiced incident response plans or standard operating procedure (SOP) for their employees to flow in the event of DDoS attack.
	\item ISPs can be helpful in mitigating the risk of DDoS attacks by implementing Best Currect Practice 38 (BCP38) \cite{bcp38} proposed by Internet Engineering Taskforce (IETF). BCP38 basically mandates that the packets with the source IP address different from the assigned address space of its originating network should be filtered out at the network ingress. This would have an immediate impact on many DDoS attacks employing reflection techniques but unfortunately, it is not widely implemented by ISPs.
	\item IoT device certification for security by national or international regulatory bodies can help promote security awareness to device manufacturers, vendors and end-users.
	\item Insurance, especially cyber insurance shall be made mandatory for commercial and private users of IoT devices to their cover 1st and 3rd party liabilities. This can be a vital tool in reducing or managing their cyber risk.
\end{itemize}

\subsection{Best Practices}
\label{sec:best}
In this section, we will discuss some mitigation steps and best practices for protection against IoT malware for the end-users. Soon after major DDoS attacks in October 2016, US-CERT of US Department of Homeland Security issued an Alert (TA16-288A) regarding the DDoS threat posed by Mirai and other IoT botnets \cite{us-cert}. They proposed different mitigation and preventive steps to protect end-users from IoT malware infection to their Internet enabled devices. 

The mitigation steps for the end-users after the IoT malware infection are as follows:

\begin{enumerate}
	\item Disconnect the device from the Internet.
	\item Reboot the device, as most IoT malware exists in temporary memory (RAM) and a reboot will clear the device from IoT malware.
	\item Ensure the default password is changed to a strong password to avoid re-infection by the IoT malware. 
	\item Update the firmware, if available. Many manufacturers of vulnerable devices provide security patches after an array of DDoS attacks in Q3 of 2016.
	\item Reconnect to the network.
\end{enumerate}

In order to prevent a IoT malware infection, end-users of the IoT devices shall follow the best practices listed below:

\begin{itemize}
	\item Change the default password to strong password. Most IoT malwares use default credentials to perform brute force attack to gain telnet access to the devices.
	\item Always update the IoT devices with security patches provided by their manufacturers as soon as they are available.
	\item Disable all ports and services on IoT devices which are not used. 
	\item Diable Universal Plug and Play on routers unless it is absolutely necessary
	\item Isolate the IoT devices on their own protected network using firewalls or using different network segmentation techniques. Don’t allow unnecessary Internet access to IoT devices.
	\item Purchase IoT devices from reputed manufacturers with good track record for producing secure devices or  responding with regular security updates.
	\item Periodically monitor the Internet firewall logs for anomaly or suspicious traffic. especially look for suspicious traffic on ports 2323 (telnet traffic) and 23/TCP (telnet traffic)
	\item End-user should be aware of the capabilities and application of the IoT devices installed. 
\end{itemize}

\subsection{Cyber insurance}
\label{sec:Cyberinsurance}
According to the Federal Office for Information Security, cyber security problems cost the german economy about 45-50 million euros per year. IT head of Volkswagen, Martin Hofmann, said in Auguest 2016 that Volkswagen had to handle 6000 cyber attacks a week. DDoS using IoT devices is becoming a growing menace and can affect anyone on the Internet including end-users, device manufacturers and vendors. In addition to deploying security controls to safeguard and mitigate risks from DDoS attacks by IoT botnets, one can consider having cyber insurance additionally to compensate economic losses incurred by any cyber incidents. Cyber insurance is an important tool for corporate and private user to manage and reduce cyber risks to their vital assets. 

\begin{figure}
\begin{center}
  \includegraphics[width=0.48\textwidth]{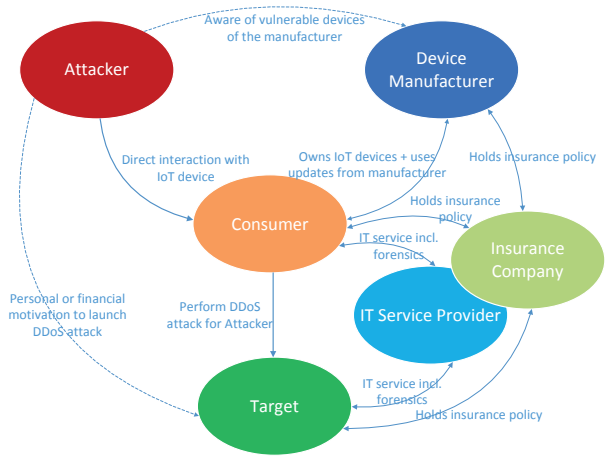}
\end{center}
\caption{Relationship between different actors involved in DDoS using IoT devices}
\label{fig:Cyberinsurance}       
\end{figure}

The actors in a cyber incident are listed below:
\begin{enumerate}
	\item Attacker - initiator of DDoS or other cyber attacks.
  \item Device manufacturers - producers of vulnerable IoT devices
	\item Consumer - owner / end user of the IoT devices
	\item Target - victim of the DDoS attack, who incurs losses of any kind due to the DDoS attack
	\item IT service provider - provides services in pre- and post cyber incidents
	\item Insurance company - provides different types of insurance to help its customers manage and reduce cyber risks.
\end{enumerate}
The relationship between different actors in a cyber incident is depicted the in the figure \ref{fig:Cyberinsurance}. The key observation is that the insurance company and IT service provider share the similar relationship with the consumers and target. The roles of an IT service provider and an insurance company in relation to consumer and target are overlapping and need a closer co-operation with each other. In fact the services offered by an insurance company is much more elaborate and includes interfacing with IT service provider for risk assessment, IT forensics for claims management. The important question for the consumers and targets is that, if they can trust the IT service providers of their choice. There has been many instances in the past where an IT service provider have installed backdoor on their client's devices for convenience of maintenance or for malicious activities. For example, many famous cyber criminal groups like Lizardsquad, Poodle Corp, vDOS offer IT services for testing the network infrastructure and also provide DDoSaaS for criminal activities. DDoS mitigation company, BackConnect provide DDoS mitigation services using very non-ethical technical means like BGP hijacking. Similarly, Paras Jha (the alleged author of Mirai botnet malware) was offering DDoS mitigation services using his company ProTraf Solutions \cite{mirai}. Therefore, it is often advisable for the consumers (and the target) to work closely with the insurance company and use the services of the IT service provider recommended by the insurance company. Close collaboration between consumers (and the target) and the insurance company can lead to a symbiosis relationship, where both the consumers (and the target) and insurance company will benefit. The recommendation from the insurance company can enhance the trust of the consumer on the IT service provider. The insurance company can use the risk assessment by the IT service provider to make meaningful cyber insurance coverage to the consumers (and the target). Crisis management, IT forensics, data recovery and system hardening services offered by the IT service provider to a target after a cyber incident will be made available faster and more reliable in accordance with the insurance coverage by the insurance company.  

Insurance can be interesting for device manufacturers, consumers and targets. Cyber insurance mostly offers following coverage elements:
\begin{enumerate}
	\item Loss or theft of data coverage (1st party)
	\item Confidentiality breach liability coverage (3rd party)
	\item Privacy breach protection coverage (1st party)
	\item Privacy liability coverage (3rd party)
	\item Payment card industry data security standard (PCI-DSS) coverage (1st party)
	\item Business interruption coverage (1st party)
	\item Cyber extortion coverage (1st party)
	\item Network security liability coverage (3rd party)
	\item Reputational risk coverage (1st party)
\end{enumerate}
with the exclusions for claims and damages for 
\begin{enumerate}
	\item physical injury to tangible property (1st party with the exception of electronic data and 3rd party)
	\item bodily injury (1st and 3rd party)
	\item product recalls (1st party), including damage to property containing an allegedly defective product (3rd party)
\end{enumerate}
Cyber insurance is not a solution for everything related to cyber incidents. It can be observed that the sort of liability exposures listed in the exclusions may be precisely the types of losses caused by a cyber attack made through the IoT. The short comings of the cyber coverage can be often addressed by negotiation with insurers during the placement process or by existing coverage under other lines of insurance, like general liability, first-party property and specialty lines coverage. However, it should be noted that there are some scenarios like outage of external networks (due to failure of power supply, telecommunication network, Internet infrastructure or others) influencing the business of the insured cannot be covered by cyber insurance.

For consumers and DDoS targets, cyber insurance covers almost all non-damage scenarios, including the costs related to crisis consulting, crisis management, notification costs, call center, credit monitoring, legal consulting, claims handling, public relations and IT services, like IT forensics, data forensics (including accounting). However, the device manufacturers would additionally need the following insurances, namely,
\begin{enumerate}
	\item Technology errors and omissions insurance (tech E\&O) - to cover 3rd party claims made by clients for inadequate work or negligent action in providing technology services or products
	\item Product liability insurance - to cover 3rd party claims due to damage to property containing an allegedly defective product
	\item product recall insurance - to cover 1st party claims due to product recalls
\end{enumerate}
For example, DVRs with hardcoded credentials manufactured by Chinese manufacturer XiongMai Technologies Technology which was used in massive DDoS attack on Dyn, was recalled in October \cite{recall}. On 5 January 2017, Federal Trade Commission (FTC) decided to sue D-Link, a Taiwanese manufacturer of networking equipments, for failing to take reasonable steps to secure their routers and IP cameras \cite{ftc}. 

\section{Conclusions}
\label{sec:conclusion}

The Internet has become ubiquitous and essential part of our lives. It has enable easy communication, more efficiency at work, connected enhanced living and accelerated innovation. At the same time, Internet has also increased the ease, viability and efficiency of launching a large scale DDoS attacks, especially using IoT devices. In 2002, the highest DDoS attack was 100 Mbps but in 2016, the highest DDoS attack is in the order of 1.1-1.5Tbps. Even though there are several peaks in the bandwidths of DDoS attacks, the average follows Moore's law (doubles every 12-24 months) which is in line with other technical developments such as CPU and storage sizes. Free availability of source code of IoT botnets like LightAidra, BASHLITE and Mirai has led to flood of many miscreants and script kiddies trying their hand at IoT malwares. Especially, IoT malware Mirai has inspired a renaissance in IoT malwares and responsible for large scale DDoS attacks for example two DDoS attacks in the order of 1.1 Tbps within a very short period of time. IoT botnet has exposed the absence of basic security in IoT devices and ignorance of best practices among IoT users. The lack of control over IoT device manufacturers, lack of security by design in the Internet and IoT infrastructure did not help much in the combat against IoT malwares. IoT botnets are evolving in sophistication and impact. If left un-checked, it can soon inflict serious impacts on critical infrastructure systems. The CDNs, DNS and ISP play an very important role to stymie DDoS by IoT botnets. The capability of the bots can be reduced by the combined efforts of device manufacturers, legislators, regulators and end-users to implement and follow basic cybersecurity and cyber-hygiene guidelines. The device manufacturers, end-users should consider cyber insurance or other lines of insurance (especially for device manufacturers) to manage and reduce cyber risks posed to their vital assets. 




\balance
\begin{table*}
\caption{List of Relevant TCP/IP Protocols}
\label{app:protocols}       
\begin{tabular}{p{1cm}p{5cm}p{10.5cm}}
\hline\noalign{\smallskip}
Protocol & Full Name & Functionality  \\
\noalign{\smallskip}\hline\noalign{\smallskip}
HTTP & Hypertext Transfer Protocol &  HTTP is an application layer protocol for distributed, collaborative, hypermedia information systems. HTTP is the foundation of data communication for the World Wide Web.\\
TR-069 & Technical Report 069 & TR-069 is a technical specification that defines an application layer protocol for remote management of end-user devices. It was published by the Broadband Forum and entitled CPE WAN Management Protocol (CWMP) \\
SOAP & Simple Object Access Protocol &  SOAP is an application layer protocol for exchanging structured information between client and web services using Extensible Markup Language (XML).\\
FTP & File Transfer Protocol &  FTP is an application layer protocol used for transfer of files between a client and server on a IP based network. \\
Telnet & Telnet &  Telnet is an application layer protocol used for bidirectional interactive text-oriented communication over the IP based network.\\
DNS & Domain Name System &  The Domain Name System is a hierarchical decentralized naming system for any devices connected to Internet or Intranet. DNS used by any network enabled devices to translate commonly used domain names of the destination server into their corresponding IP address to enable communication on a IP based networks. As a worldwide directory service, DNS is important for the normal functioning of Internet.\\
SSL  & Secure Socket Layer &  Secure Sockets Layer (SSL) is application layer cryptographic protocols that provide end-to-end communication security for the transport layer of IP network.\\
TLS  & Transport Layer Security &  Transport Layer Security (TLS) is application layer cryptographic protocols that provide end-to-end communication security for the transport layer of IP network.\\
SSH & Secure Shell &  Secure Shell is an application layer cryptographic network protocol used for operating network services securely over an unsecured network.\\
DHCP & Dynamic Host Configuration Protocol &  DHCP is an application layer protocol used for dynamically distribute network configuration parameters such as IP addresses to network devices on a IP based network\\
BGP & Border Gateway Protocol &  BGP is an application layer protocol used to exchange routing and reachability information among autonomous (AS) on the Internet.\\
NTP & Network Time Protocol &  NTP is an application layer protocol used for clock synchronization of network devices on a variable latency IP data network.\\
SNMP & Simple Network Management Protocol &  SNMP is an application layer protocol used for collecting, organizing information about the managed devices and modify them to change behavior of the devices on IP networks.\\
TCP & Transport Control Protocol &  TCP is one of the core transport layer protocol in TCP/IP protocol suite used for reliable, ordered and error free delivery of packets between two hosts communicating on IP networks.\\
UDP & User Datagram Protocol &  UDP is one of the another core transport layer protocol in TCP/IP protocol suite used for unreliable, unordered but fast delivery of packets between two hosts communicating on IP networks.\\
IP & Internet Protocol &  IP is the principle network protocol in the TCP/IP protocol suite for relaying datagrams across network boundaries solely based on the IP addresses in the datagram header.\\
ICMP & Internet Control Message Protocol &  ICMP is a network layer protocol used by network devices, like routers, to send messages and operational information indicating issues along the path of the datagram in Internet.\\
ARP & Address Resolution Protocol &  ARP is a network layer protocol used for the resolution of IP addresses into link layer addresses which are mostly Media Access Control (MAC)addresses. \\
GRE & Generic Routing Protocol &  Generic Routing Protocol is network layer tunnelling protocol used to create peer-to-peer network by establishing point-to-point connections between network nodes.\\
\noalign{\smallskip}\hline
\end{tabular}
\end{table*}
\end{document}